\documentstyle[11pt,newpasp,twoside,epsf]{article}
\def\edcomment#1{\iffalse\marginpar{\raggedright\sl#1\/}\else\relax\fi}
\reversemarginpar

\begin{document}
\title{The Jet Angular Profile and the Afterglow Light Curves}
\author{Jonathan Granot}
\affil{Institute for Advanced Study, Princeton, NJ 08540}
\author{Pawan Kumar}
\affil{Astronomy Department, University of Texas, Austin, TX 78731}
\author{Tsvi Piran}
\affil{Racah Institute for Physics, Hebrew University, Jerusalem 91904, Israel}

\begin{abstract}
We investigate how the angular structure of GRB jets 
effects the afterglow light curves at different
viewing angles, $\theta_{v}$, from the jet symmetry axis.
A numerical hydrodynamical modeling for the evolution 
of a relativistic collimated outflow, as it interacts with the 
surrounding medium, is carried out, and compared to two simple
models that make opposite and extreme assumptions for the degree 
of lateral energy transfer. The Lorentz factor, $\Gamma$, and kinetic 
energy per unit solid angle, $\epsilon$, are initially taken to be 
power laws of the angle $\theta$ from the jet axis. 
We find that the lateral velocity in the comoving frame, $v'_\theta$, 
is typically much smaller than the sound speed, $c_s$, as long as 
$\Gamma\gg 1$, and the dynamics of relativistic structured jets 
may be reasonably described by a simple analytic model where 
$\epsilon$ is independent of time, as long as $\Gamma(\theta=0)\ga$ 
a few. We perform a qualitative comparison between the resulting light 
curves and afterglow observations. This constrains the jet structure, 
and poses problems for a `universal' jet model, where all GRB jets are 
assumed to be intrinsically identical, and differ only by our viewing 
angle, $\theta_{v}$. 
\end{abstract}

\section{Introduction}

Most GRB jet models consider a uniform (or `top hat') jet, where the Lorentz 
factor, $\Gamma$, and the energy per unit solid angle, $\epsilon$, do not 
depend on the angle $\theta$ from the jet axis, within some finite well defined
opening angle, $\theta_j$, and drop sharply outside of $\theta_j$. 
The possibility that GRB jets can display an angular
structure, where $\epsilon$ and $\Gamma$ are power laws in $\theta$,
was proposed by M\'esz\'aros, Rees \& Wijers (1998). 
Recently, several different groups have analyzed afterglow observations within
the frame work of the `top hat' jet model, and have inferred a relatively
narrow distribution both for the total energy output in gamma-rays
(Frail et al. 2001) and in the initial kinetic energy of the relativistic
outflow (Panaitescu \& Kumar 2001; Piran et al. 2001). These results
may alternatively be interpreted as GRB jets having a universal structure,
which is intrinsically the same for all GRBs, and the observed
differences between different GRBs are a result of different viewing
angles, $\theta_v$, w.r.t the jet symmetry axis
(Lipunov, Postnov \& Prokhorov 2001; Rossi, Lazzati \& Rees 2002;
Zhang \& M\'esz\'aros 2002). Whereas in the `top hat' jet interpretation,
the jet break time, $t_j$, depends mainly on the initial opening angle of
the jet, $\theta_j$, in the universal `structured' jet interpretation, $t_j$
depends mainly on the viewing angle, $\theta_{v}$, and the light curve
is roughly similar to that for a `top hat' jet with $\theta_j=\theta_v$ and 
$\epsilon=\epsilon(\theta_v)$.

While the evolution of `top hat' jets and their light
curves has been widely investigated, 
much less work has been done on `structured' jets. 
Here we describe the main results of Kumar \& Granot (2003, KG hereafter) 
and Granot \& Kumar (2003, GK hereafter), and refer the 
reader to these works for more details.

\section{The Jet Dynamics \& Afterglow Light Curves}

\begin{figure}
\plotone{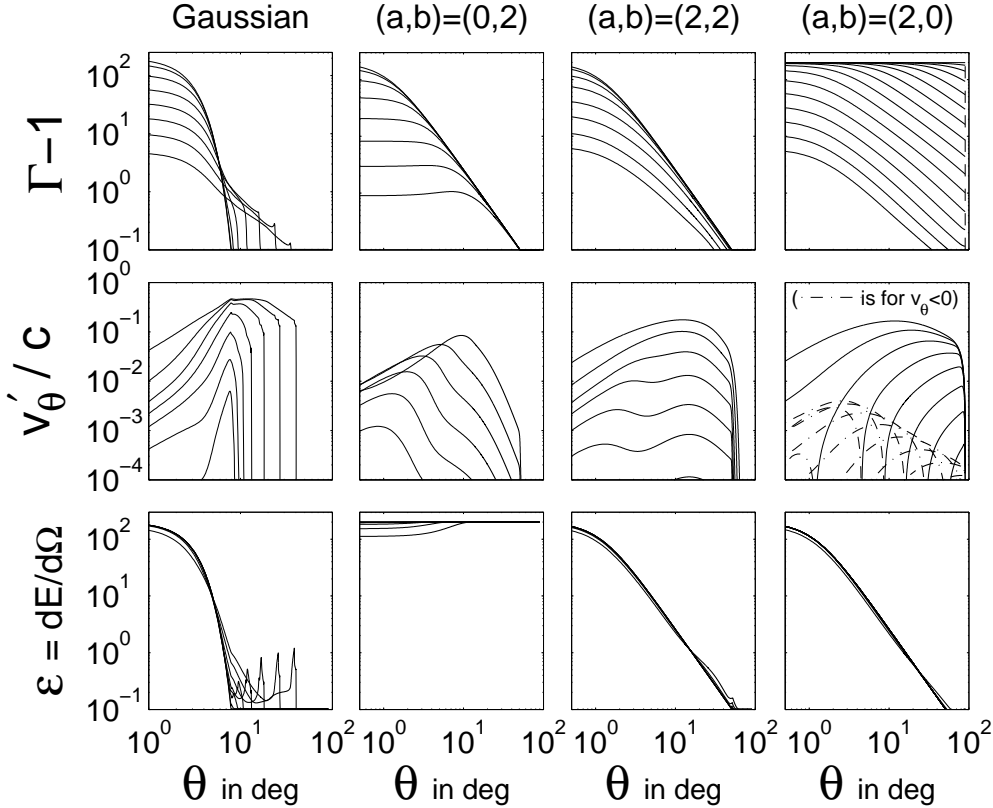}
\caption{\label{fig1}The hydrodynamic evolution of a structured jet,
for $\Gamma_0=200$, $\epsilon_0=(10^{53}\;{\rm erg})/(4\pi\;{\rm sr})$.
We show $\Gamma-1$, the lateral velocity in the local frame, 
$v'_\theta/c=\Gamma v_\theta/c$, and $\epsilon$, for an initial Gaussian
profile, and three different power law profiles where initially
$\epsilon\propto\theta^{-a}$ and $\Gamma\propto\theta^{-b}$.}
\end{figure}

We begin with a numerical hydrodynamic calculation of the evolution 
of a relativistic jet (KG). The hydrodynamic equations are reduced to 
1-D by assuming axial symmetry and integrating over the radial 
profile of the flow, thus considerably reducing the computation time.
We examined initial conditions where $\epsilon$ and $\Gamma$
are power laws in $\theta$, outside of a core angle, $\theta_c$: 
$\epsilon(\theta,t_0)=\epsilon_0\Theta^{-a}$ and
$\Gamma(\theta,t_0)=1+(\Gamma_0-1)\Theta^{-b}$, where 
$\Theta\equiv[1+(\theta/\theta_c)^2]^{1/2}$, as well as a Gaussian
profile: $[\Gamma(\theta,t_0)-1]/(\Gamma_0-1)=\epsilon(\theta,t_0)/\epsilon_0
=\exp(-\theta^2/2\theta_c^2)$. 
For simplicity, we show results only for a uniform ambient medium. 
The hydrodynamic evolution is shown in Fig. 1. 
For power law jet profiles the lateral energy transfer is 
small, and as long as $\Gamma(\theta=0)\ga\;$a few, 
$\epsilon(\theta,t)\approx\epsilon(\theta,t_0)$. For the Gaussian profile,
which is the steepest, a shock forms in the lateral direction, but most of 
the energy still remains at $\theta\la\theta_c$. The lateral velocity
in the comoving frame is found to be 
$v'_\theta\sim c/\Gamma\ll c_s\approx c/\sqrt{3}$ for $\Gamma\gg 1$
(in KG we show analytically that $v'_\theta\sim c/(\Gamma\delta\theta)$ 
where $\delta\theta$ is the angle on which $\Gamma$ or $\epsilon$ 
change significantly).

\begin{figure}
\plotone{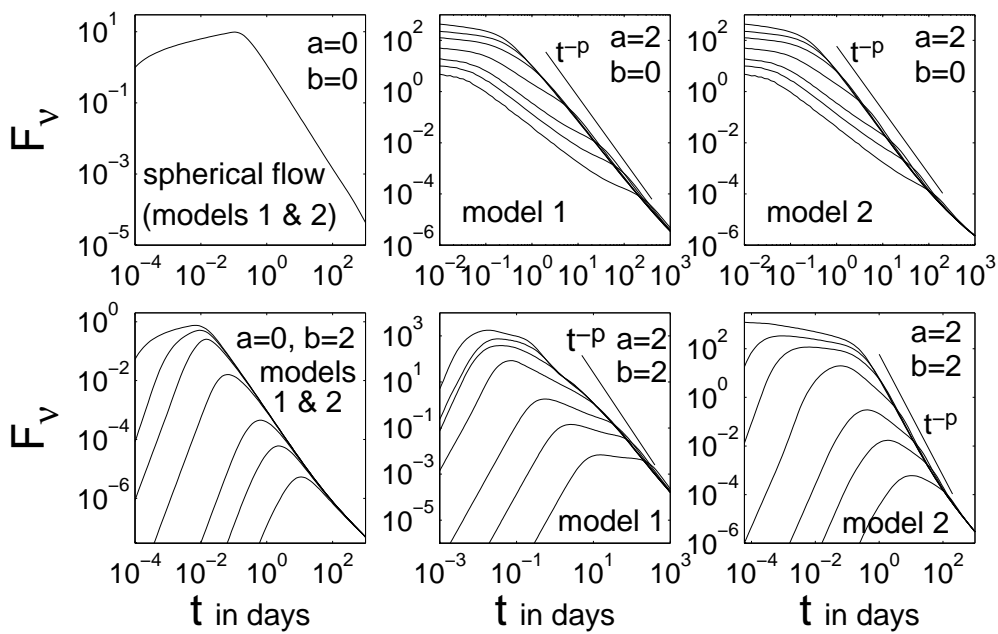}
\caption{\label{fig1}Light curves for structured jets 
(initially $\epsilon\propto\theta^{-a}$ and $\Gamma\propto\theta^{-b}$),
for models 1 and 2, in the optical
($\nu=5\times 10^{14}\;$Hz), for a jet core angle $\theta_c=0.02$,
viewing angles $\theta_{\rm obs}=0.01,0.03,0.05,0.1,0.2,0.3,0.5$,
$p=2.5$, $\epsilon_e=\epsilon_B=0.1$, $n=1\;{\rm cm}^{-3}$,
$\Gamma_0=10^3$, and $\epsilon_0$ was chosen so that the total energy of
the jet would be $10^{52}\;$erg (GK). A power law of $t^{-p}$ is added in some
of the panels, for comparison.}
\end{figure}

Now we  examine two simple models, where either: 
(1) $\epsilon(\theta,t)=\epsilon(\theta,t_0)$, or (2) $\epsilon$
is averaged over the region to which a sound wave can propagate
(this simulates the maximal lateral energy transfer that is 
consistent with causality). We assume initial power law profiles, 
as described above, and calculate the afterglow light curves for 
observers at different viewing angles, $\theta_v$, assuming 
synchrotron emission (see Fig. 2). For $(a,b)=(0,2)$ the light curve 
initially rises [before the deceleration time $t_{\rm dec}(\theta_v)$],
and there is no jet break, which is quite different from observations 
for most afterglows. For $(a,b)=(2,2),\,(2,0)$ we find a jet break
at $t_j$ when $\Gamma(\theta_v)\sim\theta_v^{-1}$. For $(a,b)=(2,2)$ 
the value, $\alpha_1$, of the temporal decay slope 
$\alpha\equiv d\log F_\nu/d\log t$ at $t<t_j$, increases with $\theta_v$,
while $\alpha_2=\alpha(t>t_j)$ decreases with $\theta_v$. This effect is 
more prominent in model 1, and appears to a lesser extent in model 2, 
but is prominent in the light curves from the numerical calculation of 
the jet dynamics (KG). This suggests that $\delta\alpha=\alpha_1-\alpha_2$
should increase with $t_j$, which is not supported 
by observations. For $(a,b)=(2,0)$, there is a flattening of the 
light curve just before the jet break, for $\theta_v\ga 3\theta_c$.
Again, this effect is larger in model 1, compared to model 2.
The light curves for the simulated jet dynamics (KG) show a reasonably 
sharp jet break at $\theta_v\ga (2-3)\theta_c$ and flattening of the 
light curve just before the jet break becomes strong at 
$\theta_v\ga(5-7)\theta_c$. This leaves a factor of $\sim 3$ in 
$\theta_v/\theta_c$ for which there is a sharp jet break not preceded 
by a flattening, as is typically observed in afterglow light curves.
Since a larger range of $\theta_v$ ($\sim\theta_j$ for `top hat' jets)
is inferred ($\sim 2-20^\circ$), this poses a problem for a universal 
jet structure.

\section{Conclusions}

We have described a numerical scheme for calculating the dynamics of 
structured relativistic jets. The lateral velocity in the local frame 
is found to be smaller than the sound speed, except for very sharp
jet angular profiles. This causes the energy per unit solid angle, $\epsilon$,
to remain close to its initial value in the first few days, suggesting
that a simple model (model 1) where $\epsilon$ is constant in time, and 
each segment of the jet evolves as if it were part of a spherical flow, 
provides a good approximation for the jet dynamics and light curves for 
$t\la\;$a few days. 

A universal jet profile, where all GRB jets are intrinsically identical,
and only our viewing angle changes between different GRBs,
has difficulty explaining all the afterglow light curves observed to date.
This was demonstrated for $\epsilon(t_0)\propto\theta^{-2}$, which is 
needed to reproduce the recent results of a roughly constant energy 
in GRB jets. Furthermore, we find that the jet break for structured jets
in an external density $\propto r^{-2}$ are much smoother compared to a 
uniform density, in agreement with the result of Kumar \& Panaitescu (2000)
for a `top hat' jet. Finally, the light curves for a Gaussian initial 
jet profile are found to be similar to those for a `top hat' jet, and 
(for $\theta_v\la\theta_c$) produce sharp jet breaks, 
that are compatible with afterglow observations.

\end{document}